\pdfoutput=1

\documentclass[12pt,a4paper]{article}

\usepackage{ifthen} 
\newboolean{pdflatex}
\setboolean{pdflatex}{true} 

\newboolean{articletitles}
\setboolean{articletitles}{true} 

\newboolean{uprightparticles}
\setboolean{uprightparticles}{false} 

\newboolean{inbibliography}
\setboolean{inbibliography}{false} 

\usepackage{microtype}
\usepackage{lineno}  
\usepackage{xspace} 
\usepackage{caption} 

\usepackage{graphicx}  
\usepackage{color}
\usepackage{colortbl}
\graphicspath{{./figs/}} 

\usepackage{amsmath} 
\usepackage{amssymb}
\usepackage{amsfonts}
\usepackage{upgreek} 

\newcommand*\patchAmsMathEnvironmentForLineno[1]{%
\expandafter\let\csname old#1\expandafter\endcsname\csname #1\endcsname
\expandafter\let\csname oldend#1\expandafter\endcsname\csname
end#1\endcsname
 \renewenvironment{#1}%
   {\linenomath\csname old#1\endcsname}%
   {\csname oldend#1\endcsname\endlinenomath}%
}
\newcommand*\patchBothAmsMathEnvironmentsForLineno[1]{%
  \patchAmsMathEnvironmentForLineno{#1}%
  \patchAmsMathEnvironmentForLineno{#1*}%
}
\AtBeginDocument{%
\patchBothAmsMathEnvironmentsForLineno{equation}%
\patchBothAmsMathEnvironmentsForLineno{align}%
\patchBothAmsMathEnvironmentsForLineno{flalign}%
\patchBothAmsMathEnvironmentsForLineno{alignat}%
\patchBothAmsMathEnvironmentsForLineno{gather}%
\patchBothAmsMathEnvironmentsForLineno{multline}%
\patchBothAmsMathEnvironmentsForLineno{eqnarray}%
}

\def\lhcb {\mbox{LHCb}\xspace}

\ifthenelse{\boolean{uprightparticles}}%
{

 \def\Ppi         {\ensuremath{\uppi}\xspace}

 \def\Ppsi        {\ensuremath{\uppsi}\xspace}

 \def\PDelta      {\ensuremath{\Delta}\xspace}                 
 \def\PXi      {\ensuremath{\Xi}\xspace}                 
 \def\PLambda      {\ensuremath{\Lambda}\xspace}                 
 \def\PSigma      {\ensuremath{\Sigma}\xspace}                 
 \def\POmega      {\ensuremath{\Omega}\xspace}                 
 \def\PUpsilon      {\ensuremath{\Upsilon}\xspace}                 
 
 \def\PB      {\ensuremath{\mathrm{B}}\xspace}                 
                  
 \def\PD      {\ensuremath{\mathrm{D}}\xspace}

 \def\PJ      {\ensuremath{\mathrm{J}}\xspace}                 
 \def\PK      {\ensuremath{\mathrm{K}}\xspace}

 \def\Pb      {\ensuremath{\mathrm{b}}\xspace}                 
 \def\Pc      {\ensuremath{\mathrm{c}}\xspace}

 \def\Pi      {\ensuremath{\mathrm{i}}\xspace}

 \def\Pp      {\ensuremath{\mathrm{p}}\xspace}

}
{

 \def\Ppi         {\ensuremath{\pi}\xspace}

 \def\Ppsi        {\ensuremath{\psi}\xspace}                 
                  
 \mathchardef\PDelta="7101
 \mathchardef\PXi="7104
 \mathchardef\PLambda="7103
 \mathchardef\PSigma="7106
 \mathchardef\POmega="710A
 \mathchardef\PUpsilon="7107
                  
 \def\PB      {\ensuremath{B}\xspace}                 
                  
 \def\PD      {\ensuremath{D}\xspace}

 \def\PJ      {\ensuremath{J}\xspace}                 
 \def\PK      {\ensuremath{K}\xspace}

 \def\Pb      {\ensuremath{b}\xspace}                 
 \def\Pc      {\ensuremath{c}\xspace}

 \def\Pi      {\ensuremath{i}\xspace}

 \def\Pp      {\ensuremath{p}\xspace}

}

\makeatletter
\ifcase \@ptsize \relax
  \newcommand{\miniscule}{\@setfontsize\miniscule{4}{5}}
\or
  \newcommand{\miniscule}{\@setfontsize\miniscule{5}{6}}
\or
  \newcommand{\miniscule}{\@setfontsize\miniscule{5}{6}}
\fi
\makeatother

\DeclareRobustCommand{\optbar}[1]{\shortstack{{\miniscule (\rule[.5ex]{1.25em}{.18mm})}
  \\ [-.7ex] $#1$}}

\def\cquark    {{\ensuremath{\Pc}}\xspace}

\def\bquark    {{\ensuremath{\Pb}}\xspace}

\def\pion   {{\ensuremath{\Ppi}}\xspace}

\def\pip    {{\ensuremath{\pion^+}}\xspace}

\def\kaon    {{\ensuremath{\PK}}\xspace}
  \def\Kbar    {{\kern 0.2em\overline{\kern -0.2em \PK}{}}\xspace}

\def\KorKbar    {\kern 0.18em\optbar{\kern -0.18em K}{}\xspace}

\def\Kp      {{\ensuremath{\kaon^+}}\xspace}
\def\Km      {{\ensuremath{\kaon^-}}\xspace}

  \def\Dbar    {{\kern 0.2em\overline{\kern -0.2em \PD}{}}\xspace}

\def\DorDbar    {\kern 0.18em\optbar{\kern -0.18em D}{}\xspace}

\def\B       {{\ensuremath{\PB}}\xspace}
\def\Bbar    {{\ensuremath{\kern 0.18em\overline{\kern -0.18em \PB}{}}}\xspace}

\def\BorBbar    {\kern 0.18em\optbar{\kern -0.18em B}{}\xspace}

\def\Bu      {{\ensuremath{\B^+}}\xspace}

\def\Bp      {{\ensuremath{\Bu}}\xspace}

\def\Bpm     {{\ensuremath{\B^\pm}}\xspace}

\def\Bc      {{\ensuremath{\B_\cquark^+}}\xspace}

\def\Bcpm    {{\ensuremath{\B_\cquark^\pm}}\xspace}

\def\jpsi     {{\ensuremath{{\PJ\mskip -3mu/\mskip -2mu\Ppsi\mskip 2mu}}}\xspace}

  \def\Y#1S{\ensuremath{\PUpsilon{(#1S)}}\xspace}

\def\proton      {{\ensuremath{\Pp}}\xspace}
\def\antiproton  {{\ensuremath{\overline \proton}}\xspace}

\def\Lz          {{\ensuremath{\PLambda}}\xspace}
\def\Lbar        {{\ensuremath{\kern 0.1em\overline{\kern -0.1em\PLambda}}}\xspace}
\def\LorLbar    {\kern 0.18em\optbar{\kern -0.18em \PLambda}{}\xspace}

\def\Lc      {{\ensuremath{\Lz^+_\cquark}}\xspace}

\def\to                 {\ensuremath{\rightarrow}\xspace}

\def\AT#1     {\ensuremath{A_{\mathrm{T}}^{#1}}\xspace}           

\def\C#1      {\ensuremath{\mathcal{C}_{#1}}\xspace}                       
\def\Cp#1     {\ensuremath{\mathcal{C}_{#1}^{'}}\xspace}                    
\def\Ceff#1   {\ensuremath{\mathcal{C}_{#1}^{\mathrm{(eff)}}}\xspace}        
\def\Cpeff#1  {\ensuremath{\mathcal{C}_{#1}^{'\mathrm{(eff)}}}\xspace}       
\def\Ope#1    {\ensuremath{\mathcal{O}_{#1}}\xspace}                       
\def\Opep#1   {\ensuremath{\mathcal{O}_{#1}^{'}}\xspace}                    

\newcommand{\tev}{\ifthenelse{\boolean{inbibliography}}{\ensuremath{~T\kern -0.05em eV}\xspace}{\ensuremath{\mathrm{\,Te\kern -0.1em V}}}\xspace}
\newcommand{\gev}{\ensuremath{\mathrm{\,Ge\kern -0.1em V}}\xspace}
\newcommand{\mev}{\ensuremath{\mathrm{\,Me\kern -0.1em V}}\xspace}
\newcommand{\kev}{\ensuremath{\mathrm{\,ke\kern -0.1em V}}\xspace}
\newcommand{\ev}{\ensuremath{\mathrm{\,e\kern -0.1em V}}\xspace}
\newcommand{\gevc}{\ensuremath{{\mathrm{\,Ge\kern -0.1em V\!/}c}}\xspace}
\newcommand{\mevc}{\ensuremath{{\mathrm{\,Me\kern -0.1em V\!/}c}}\xspace}
\newcommand{\gevcc}{\ensuremath{{\mathrm{\,Ge\kern -0.1em V\!/}c^2}}\xspace}
\newcommand{\gevgevcccc}{\ensuremath{{\mathrm{\,Ge\kern -0.1em V^2\!/}c^4}}\xspace}
\newcommand{\mevcc}{\ensuremath{{\mathrm{\,Me\kern -0.1em V\!/}c^2}}\xspace}

\def\mum  {\ensuremath{{\,\upmu\rm m}}\xspace}

\def\invfb   {\ensuremath{\mbox{\,fb}^{-1}}\xspace}

\def\ps   {\ensuremath{{\rm \,ps}}\xspace}

\def\gsim{{~\raise.15em\hbox{$>$}\kern-.85em
          \lower.35em\hbox{$\sim$}~}\xspace}
\def\lsim{{~\raise.15em\hbox{$<$}\kern-.85em
          \lower.35em\hbox{$\sim$}~}\xspace}

\def\ptot       {\mbox{$p$}\xspace}
\def\pt         {\mbox{$p_{\rm T}$}\xspace}

\def\bcvegpy    {\mbox{\textsc{Bcvegpy}}\xspace}

\def\evtgen     {\mbox{\textsc{EvtGen}}\xspace}

\def\geant      {\mbox{\textsc{Geant4}}\xspace}

\def\photos     {\mbox{\textsc{Photos}}\xspace}

\def\pythia     {\mbox{\textsc{Pythia}}\xspace}

\def\tell1  {TELL1\xspace}
\def\ukl1   {UKL1\xspace}

\newcommand{\eg}{\mbox{\itshape e.g.}\xspace}

\newcommand{\vs}{\mbox{\itshape vs.}\xspace}

\usepackage{cite} 
\usepackage{mciteplus}
\usepackage{hyperref}    
\usepackage[all]{hypcap} 

\def\pppi {\ensuremath{\Bpm \to p \bar{p} \pi^{\pm}}\xspace}

\def\pppiplus {\ensuremath{\Bp \to p \overline{p} \pi^{+}}\xspace}

\def\pppipluscomb {\ensuremath{p \bar{p} \pi^{+}}\xspace}

\def\bcpppi {\ensuremath{\Bcpm \to p \overline{p} \pi^{\pm}}\xspace}

\def\bcpppiplus {\ensuremath{\Bc \to p \overline{p} \pi^{+}}\xspace}

\begin{document}

\renewcommand{\thefootnote}{\fnsymbol{footnote}}
\setcounter{footnote}{1}


\begin{titlepage}
\pagenumbering{roman}

\vspace*{-1.5cm}
\centerline{\large EUROPEAN ORGANIZATION FOR NUCLEAR RESEARCH (CERN)}
\vspace*{1.5cm}
\hspace*{-0.5cm}
\begin{tabular*}{\linewidth}{lc@{\extracolsep{\fill}}r}
\ifthenelse{\boolean{pdflatex}}
{\vspace*{-2.7cm}\mbox{\!\!\!\includegraphics[width=.14\textwidth]{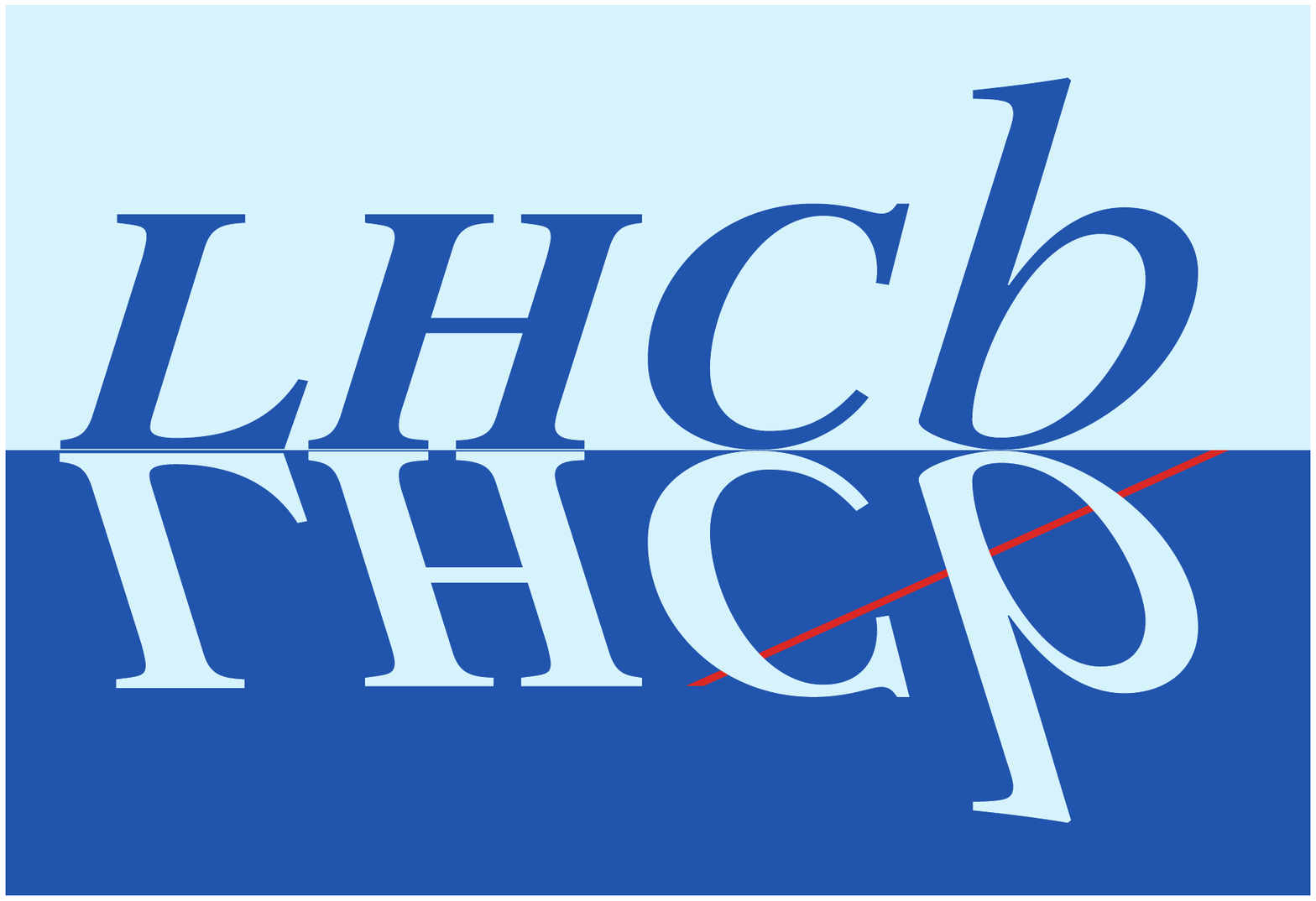}} & &}%
{\vspace*{-1.2cm}\mbox{\!\!\!\includegraphics[width=.12\textwidth]{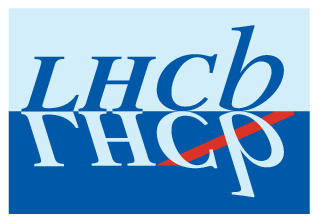}} & &}%
\\
 & & CERN-EP-2016-072 \\  
 & & LHCb-PAPER-2016-001 \\  
 & & March 22, 2016 \\ 
\end{tabular*}

\vspace*{2.0cm}

{\bf\boldmath\huge
\begin{center}
  Search for \Bc decays to the $\proton\antiproton\pip$ final state
\end{center}
}

\vspace*{1.0cm}

\begin{center}
The LHCb collaboration\footnote{Authors are listed at the end of this article.}
\end{center}

\vspace{\fill}

\begin{abstract}
  \noindent
A search for the decays of the \Bc meson to \pppipluscomb is performed for the first time using a data sample corresponding to an integrated luminosity of 3.0\invfb collected by the \lhcb experiment in $pp$ collisions at centre-of-mass energies of $7$ and $8$\tev. No signal is found and an upper limit,  at 95\% confidence level, is set, $\frac{f_c}{f_u}\times\mathcal{B}(\bcpppiplus)<3.6\times10^{-8}$ in the kinematic region $m(\proton\antiproton)<2.85\gevcc$, $\pt(B)<20\gevc$ and $2.0<y(B)<4.5$, where $\mathcal{B}$ is the branching fraction and $f_c$ ($f_u$) is the fragmentation fraction of the $b$ quark into a \Bc (\Bp) meson.

\end{abstract}

\vspace*{1.0cm}

\begin{center}
  Published in Phys.~Lett.~B 759, Pages 313-321 (2016)
\end{center}

\vspace{\fill}

{\footnotesize 
\centerline{\copyright~CERN on behalf of the \lhcb collaboration, license \href{http://creativecommons.org/licenses/by/4.0/}{CC-BY-4.0}.}}
\vspace*{2mm}

\end{titlepage}


\newpage
\setcounter{page}{2}
\mbox{~}


\cleardoublepage


\renewcommand{\thefootnote}{\arabic{footnote}}
\setcounter{footnote}{0}



\pagestyle{plain} 
\setcounter{page}{1}
\pagenumbering{arabic}


%

\section{Introduction}
The decays of the \Bc meson have the special feature of proceeding through either of its valence quarks $\overline b$ or $c$, or via the annihilation of the two.\footnote{Charge-conjugation is implied throughout the paper.} In the Standard Model, the decays with a $b$-quark transition and no charm particle in the final state can proceed only via  $\overline b c\to W^+\to u\overline q$ ($q=d,s$) annihilation, with an amplitude proportional to the product of CKM matrix elements $V_{cb}V_{uq}^*$. Cabibbo suppression $|V_{us}/V_{ud}|\sim 0.2$ implies that final states without strangeness dominate. Calculations involving two-body and quasi two-body modes predict branching fractions in the range $10^{-8}-10^{-6}$\cite{LHCb_BcPred,XiaoLiuLu,XiaoAndLiu}. Due to their rareness, the observation of these processes is an experimental challenge. On the other hand, any observation could probe other types of $\overline b c$ annihilations involving particles beyond the Standard Model, such as a mediating charged Higgs boson (see \eg Refs.~\cite{Hou_Lepton,Extended_Higgs}). 

The decays of \Bc mesons to three light charged hadrons provide a good way to study such processes. These include fully mesonic $h^{\prime+}h^{\prime-}h^+$ states or states containing a proton-antiproton pair and a light hadron, $\proton\antiproton h^+$ ($h,~h^{\prime}=\pi,~K$). 
In this study, the primary focus is on $\Bc\to\proton\antiproton \pi^+$ decays in the region below the charmonium threshold, taken to be $m(\proton\antiproton)<2.85\gevcc$, where the only contribution arises from the annihilation process. The $b\to c$ transitions, leading to $\Bc\to [c\overline c](\to \proton\antiproton)h^+$ charmonium modes, are also considered. An analysis is performed to examine these different contributions in the $\proton\antiproton \pi^+$ phase space. The $\Bp\to\proton\antiproton \pi^+$ decays in the region $m(\proton\antiproton)<2.85\gevcc$ are used as a normalization mode to derive the quantity
\begin{equation}
R_p \equiv \frac{f_c}{f_u}\times\mathcal{B}(\bcpppiplus),
\end{equation}
where $\mathcal{B}$ is the branching fraction and $f_c~(f_u)$ represents the fragmentation fraction of the $b$ quark into the $\Bc~(\Bp)$ meson. The quantity $R_p$ is measured in the fiducial region $\pt(B)<20\gevc$ and $2.0<y(B)<4.5$, where $y$ denotes the rapidity and \pt is the component of the momentum transverse to the beam. The full Run 1 (years 2011 and 2012) data sample is exploited, representing 1.0 and 2.0 \invfb of integrated luminosity at $7$ and $8$ \tev centre-of-mass energies in $pp$ collisions, respectively.

\section{Detector and simulation}

The \lhcb detector~\cite{Alves:2008zz,LHCb-DP-2014-002} is a single-arm forward
spectrometer covering the \mbox{pseudorapidity} range $2<\eta <5$, designed for the study of particles containing \bquark or \cquark quarks. The detector includes a high-precision tracking system consisting of a silicon-strip vertex detector surrounding the $pp$ interaction region, a large-area silicon-strip detector located upstream of a dipole magnet with a bending power of about $4{\mathrm{\,Tm}}$, and three stations of silicon-strip detectors and straw drift tubes placed downstream of the magnet. The tracking system provides a measurement of momentum, \ptot, of charged particles with a relative uncertainty that varies from 0.5\% at low momentum to 1.0\% at 200\gevc. The minimum distance of a track to a primary vertex (PV), the impact parameter (IP), is measured with a resolution of $(15+29/\pt)\mum$, where \pt is in\,\gevc.
Different types of charged hadrons are distinguished using information from two ring-imaging Cherenkov detectors. 
Photons, electrons and hadrons are identified by a calorimeter system consisting of scintillating-pad and preshower detectors, an electromagnetic calorimeter and a hadronic calorimeter. Muons are identified by a system composed of alternating layers of iron and multiwire proportional chambers. 

The online event selection is performed by a trigger~\cite{LHCb-DP-2012-004}, which consists of a hardware stage, based on information from the calorimeter and muon systems, followed by a software stage, which applies a full event reconstruction. At the hardware trigger stage, events are required to have a muon with high \pt or a  hadron, photon or electron with high transverse energy in the calorimeters. For hadrons,  the transverse energy threshold is 3.5\gev. The software trigger requires a two-, three- or four-track secondary vertex with a significant displacement from the primary $pp$ interaction vertices. At least one charged particle must have a transverse momentum $\pt > 1.7\gevc$ and be inconsistent with originating from a PV. A multivariate algorithm~\cite{BBDT} is used for the identification of secondary vertices consistent with the decay of a \bquark hadron.

The analysis uses simulated events generated by \pythia~8.1~\cite{Sjostrand:2007gs} and \bcvegpy \cite{BcVegPy} for the production of \Bp and \Bc mesons, respectively, with a specific \lhcb configuration~\cite{LHCb-PROC-2010-056}.  Decays of hadronic particles are described by \evtgen~\cite{Lange:2001uf}, in which final-state radiation is generated using \photos~\cite{Golonka:2005pn}. The interaction of the generated particles with the detector, and its response, are implemented using the \geant toolkit~\cite{Allison:2006ve,*Agostinelli:2002hh} as described in Ref.~\cite{LHCb-PROC-2011-006}.

\section{Reconstruction and selection of candidates}
Three charged particles are combined to form $B_{(c)}^+\to\proton\antiproton\pip$ decay candidates, which are associated to the closest PV. A loose preselection is performed on tracking quality, $p$, \pt and IP of the \Bc and its daughters, and \Bc candidate flight distance. At this stage, two windows of the invariant mass of the $p\bar{p}\pi^+$ system are retained: the \Bu region, $[5.1,5.5]\gevcc$, and the \Bc region, $[6.0,6.5]\gevcc$. Since the production fractions of different $B$ species are involved, a fiducial requirement is imposed to define the kinematic region for the measurement, $\pt(B)<20\gevc$ and $2.0<y(B)<4.5$ \cite{LHCb-PAPER-2014-050}.

\begin{figure*}[t]
\begin{center}
\includegraphics[width=0.5\textwidth]{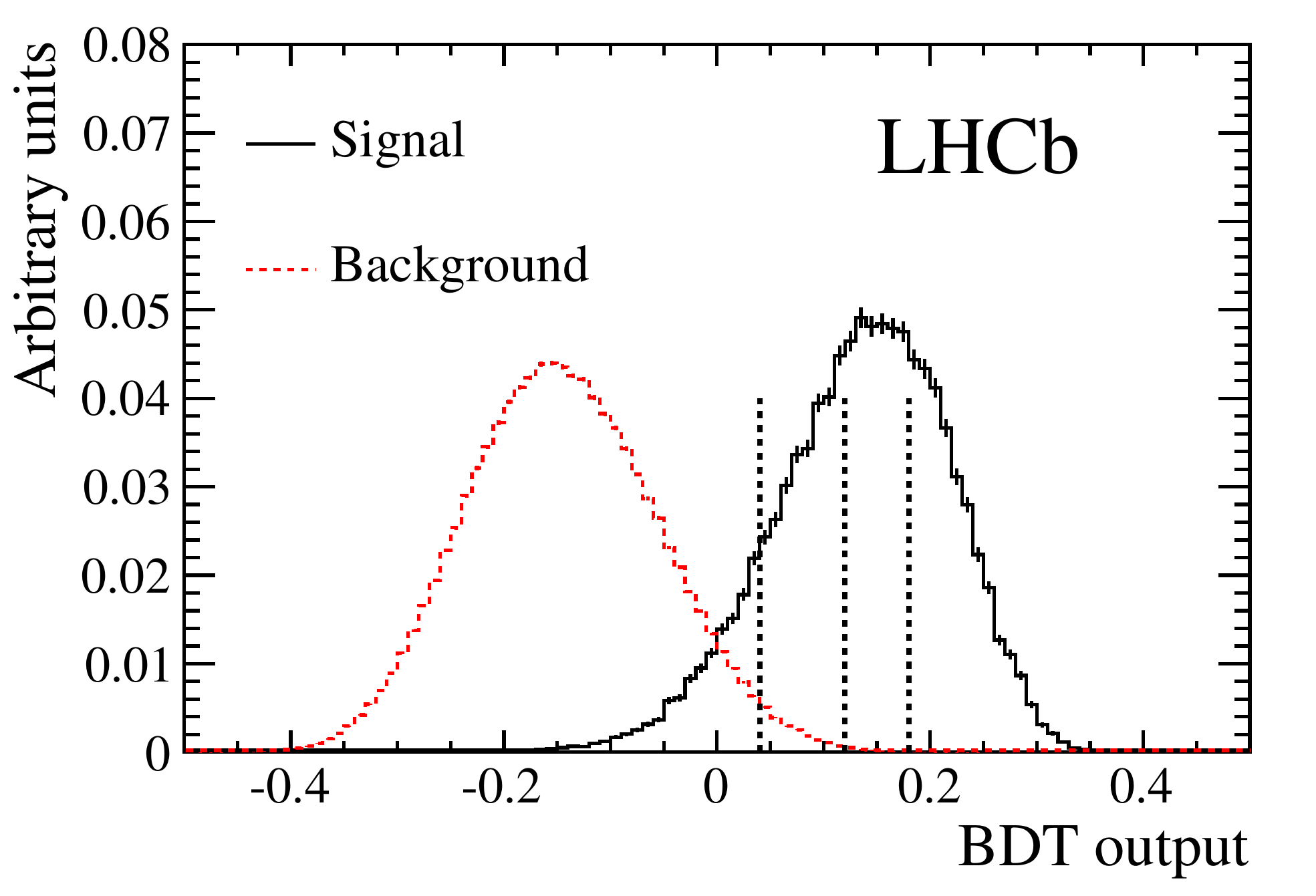}
\end{center}
\caption{Distributions of BDT output for the \bcpppiplus signal and the background. The vertical dashed lines indicate the lower limits of the three regions in which the signal is determined.}
\label{Fig:BDT_output}
\end{figure*}

Further discrimination between signal and background is provided by a multivariate analysis using a boosted decision tree (BDT) classifier\cite{BDT_theory}. Input quantities include kinematic and topological variables related to the $\Bc$ candidates and the individual daughter particles. The momentum, vertex and flight distance of the $\Bc$ candidate are exploited, as are track fit quality criteria, IP and momentum information of the final-state particles. The BDT is trained using simulated signal events, and data events from the sidebands of the $p\bar{p}\pi^+$ invariant mass $[6.0,6.15]\gevcc$ and $[6.35,6.5]\gevcc$, which represent the background. To check for training biases, the signal and background samples are split into two subsamples for training and testing of the BDT output. Figure \ref{Fig:BDT_output} shows the distribution of the BDT output for signal and background.

Particle identification (PID) requirements are applied to reduce the combinatorial background and suppress the cross-feed of $p\antiproton \Kp$ final states in the $p\antiproton \pip$ spectrum, due to the kaon being misidentified as a pion. The BDT and PID requirements are optimized jointly in order to maximize the sensitivity to very small event yields. 
The \Bc signal yield is determined from a simultaneous fit in three bins of the BDT output $X$, $0.04<X<0.12$, $0.12<X<0.18$ and $X>0.18$, each having the same expected yield (dashed lines in Fig.~\ref{Fig:BDT_output}). From simulated pseudoexperiments, this method is shown to be more sensitive than a single fit to the highest signal purity region, $X>0.18$.
The normalization channel \pppiplus undergoes the same PID and BDT selection, but its yield is determined without binning in BDT output.

\section{Fits to the data}
 Signal and background yields are obtained using unbinned extended maximum likelihood fits to the distribution of the invariant mass of the $p\antiproton \pi^+$ combinations. The \bcpppiplus and \pppiplus signals are both modelled by the sum of two Crystal Ball functions \cite{CB_shape} with a common mean. For \bcpppiplus, all the shape parameters are fixed to the values obtained in the simulation while for \pppiplus, the mean and the core width are allowed to float.
A Fermi function accounts for a possible partially reconstructed component from $\Bc\to p\antiproton \rho^+$ ($\Bp\to p\antiproton \rho^+$) decays, where a neutral pion from the $\rho^+$ is not reconstructed resulting in a $p\antiproton \pip$ invariant mass below the nominal $\Bc$ ($\Bp$) mass. An asymmetric Gaussian function with power law tails is used to model a possible $p\antiproton \Kp$ cross-feed, and its contribution is found to be negligible. The combinatorial background is modelled by an exponential function. Except for this last category, all the parameters of the background components are fixed to the values obtained in simulations.

\begin{figure*}[t]
\begin{center}
\includegraphics[width=0.45\textwidth]{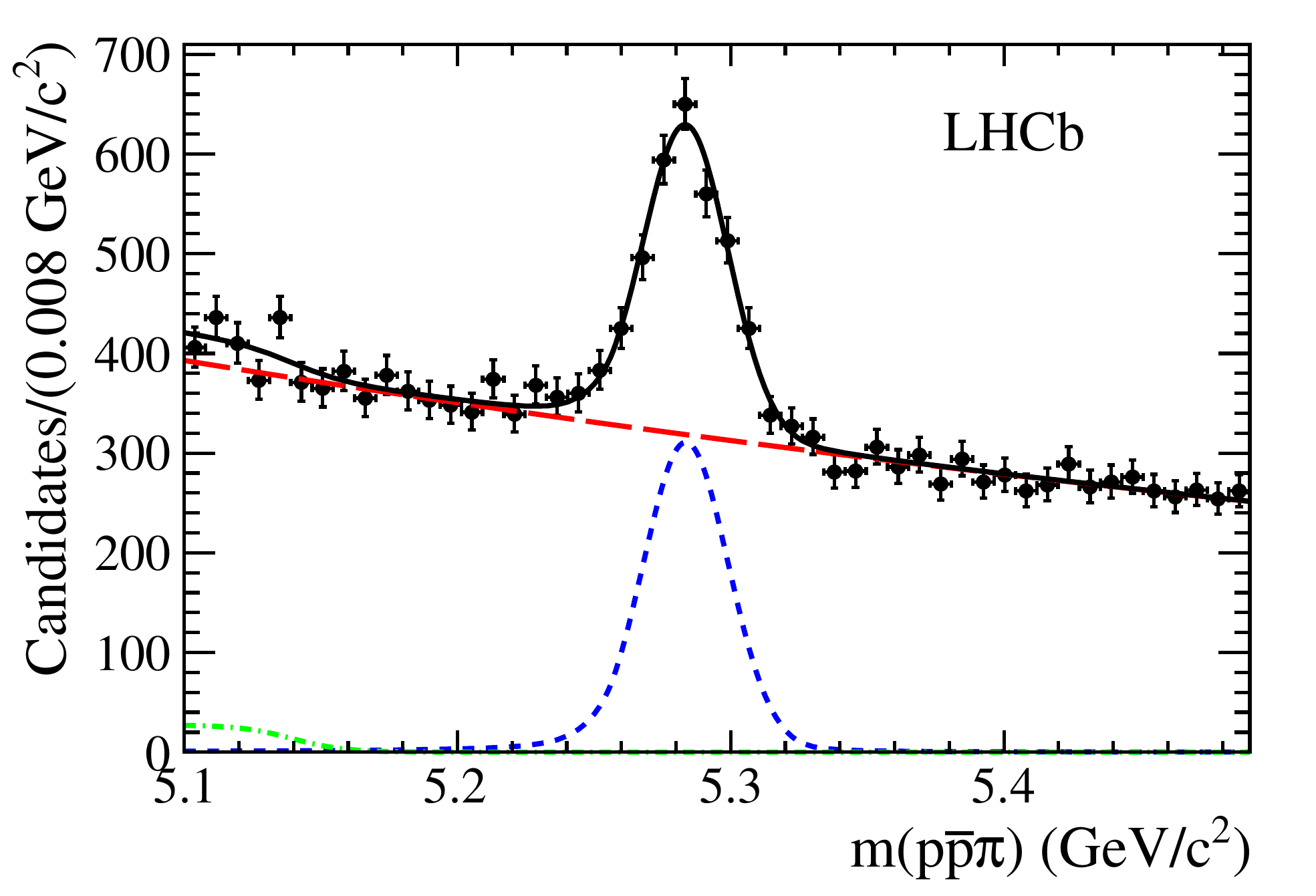}
\includegraphics[width=0.45\textwidth]{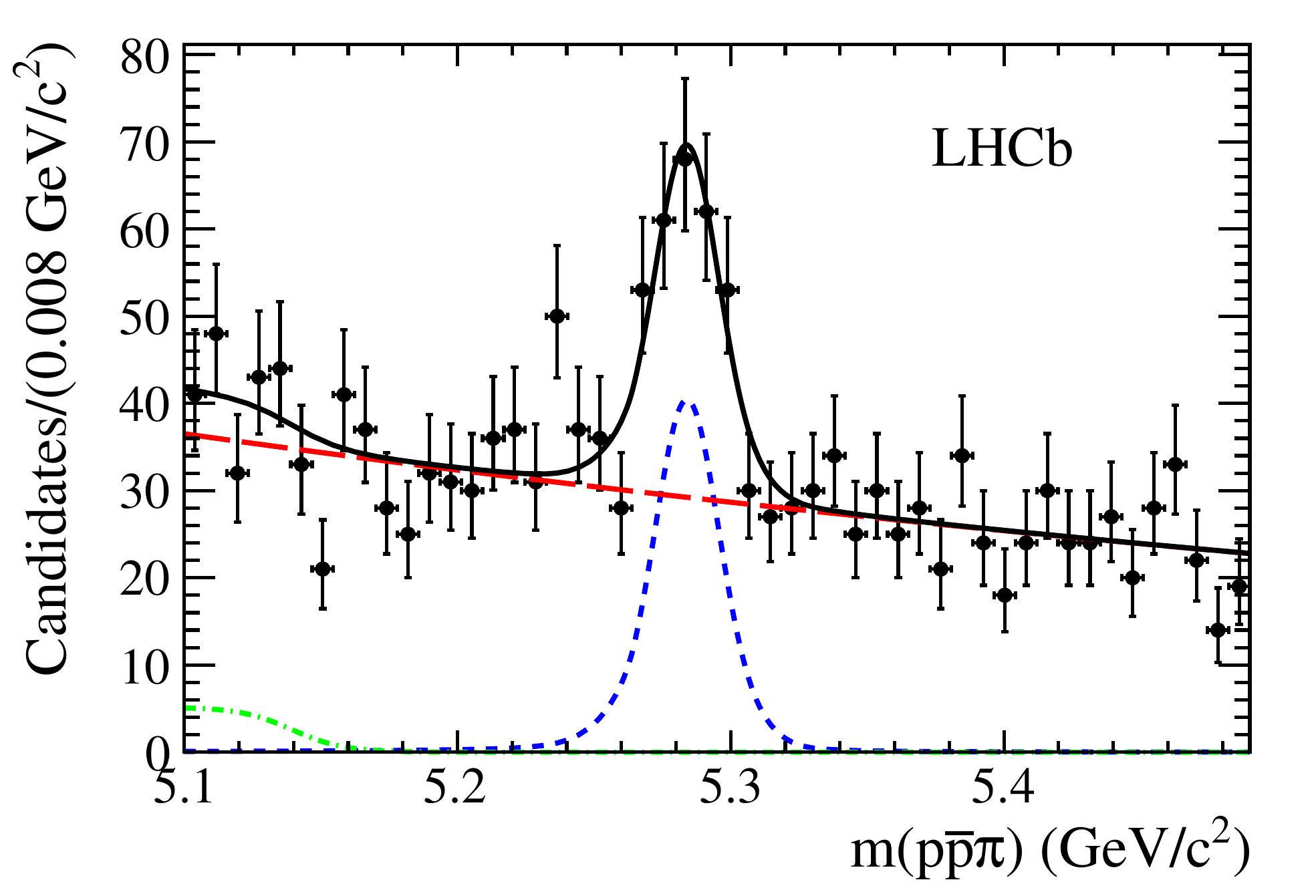}
\end{center}
\caption{Fits to the $\proton\antiproton\pip$ invariant mass in the \Bu region, for (left) $m(\proton\antiproton)<2.85\gevcc$ and (right) $2.85<m(\proton\antiproton)<3.15\gevcc$. The blue dashed, red long-dashed and green dotted-dashed lines represent the signal, combinatorial background and partially reconstructed background components, respectively. The error bars show 68\% Poisson confidence level intervals.}
\label{Fig:BuFits}
\end{figure*}

Figure \ref{Fig:BuFits} shows the result of the fits in the \Bp region. For the region of interest, $m(\proton\antiproton)<2.85\gevcc$, the yield is $N(\pppiplus)=1644\pm83$, where only the statistical uncertainty is quoted. The fit to the region $2.85<m(\proton\antiproton)<3.15\gevcc$, which includes the $\Bp\to\jpsi(\proton\antiproton)\pip$ signal, shows the yield suppression in this region as observed in Ref.~\cite{LHCb-PAPER-2014-034}.

\begin{figure*}[t]
\begin{center}
\includegraphics[width=0.45\textwidth]{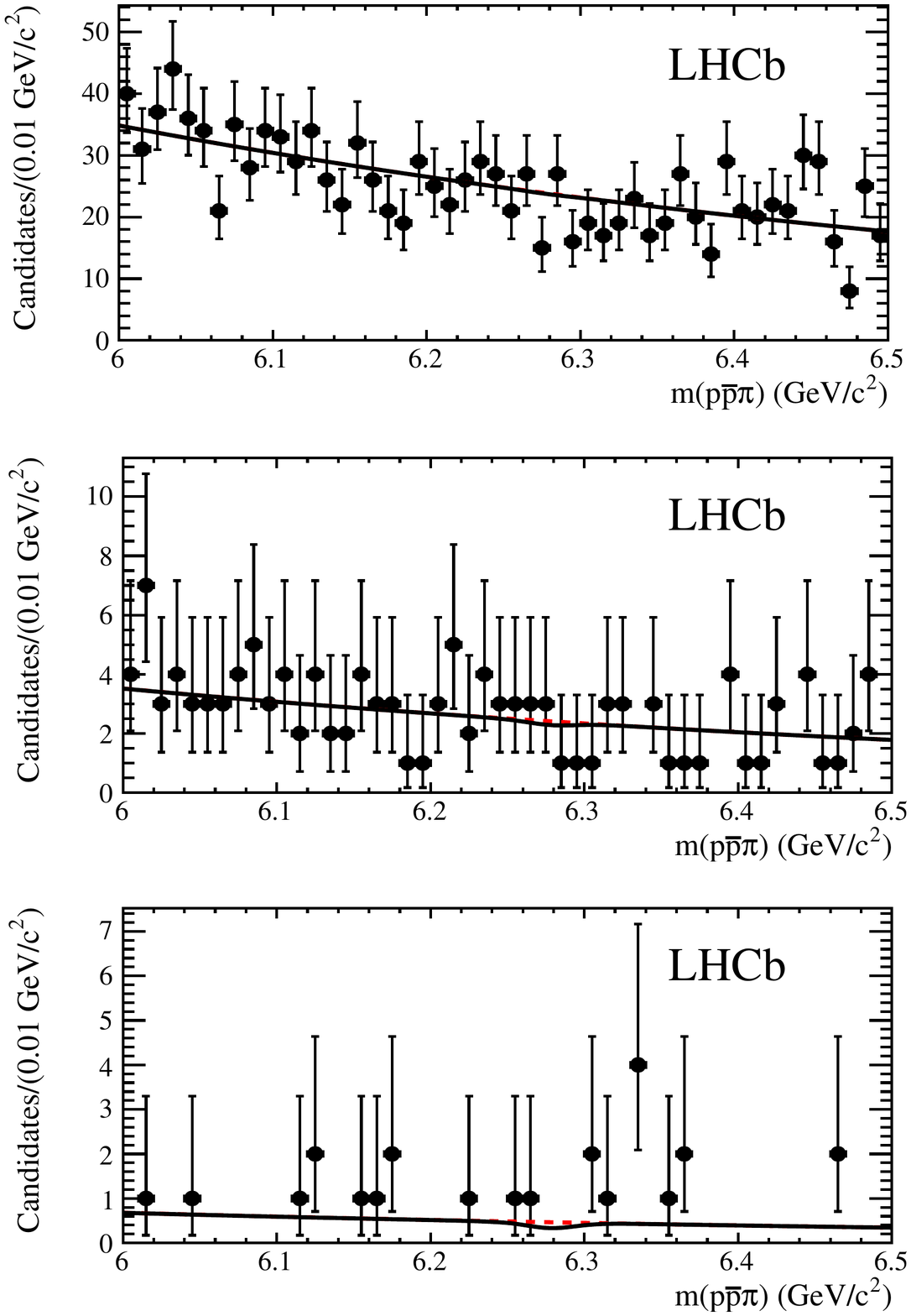}
\includegraphics[width=0.45\textwidth]{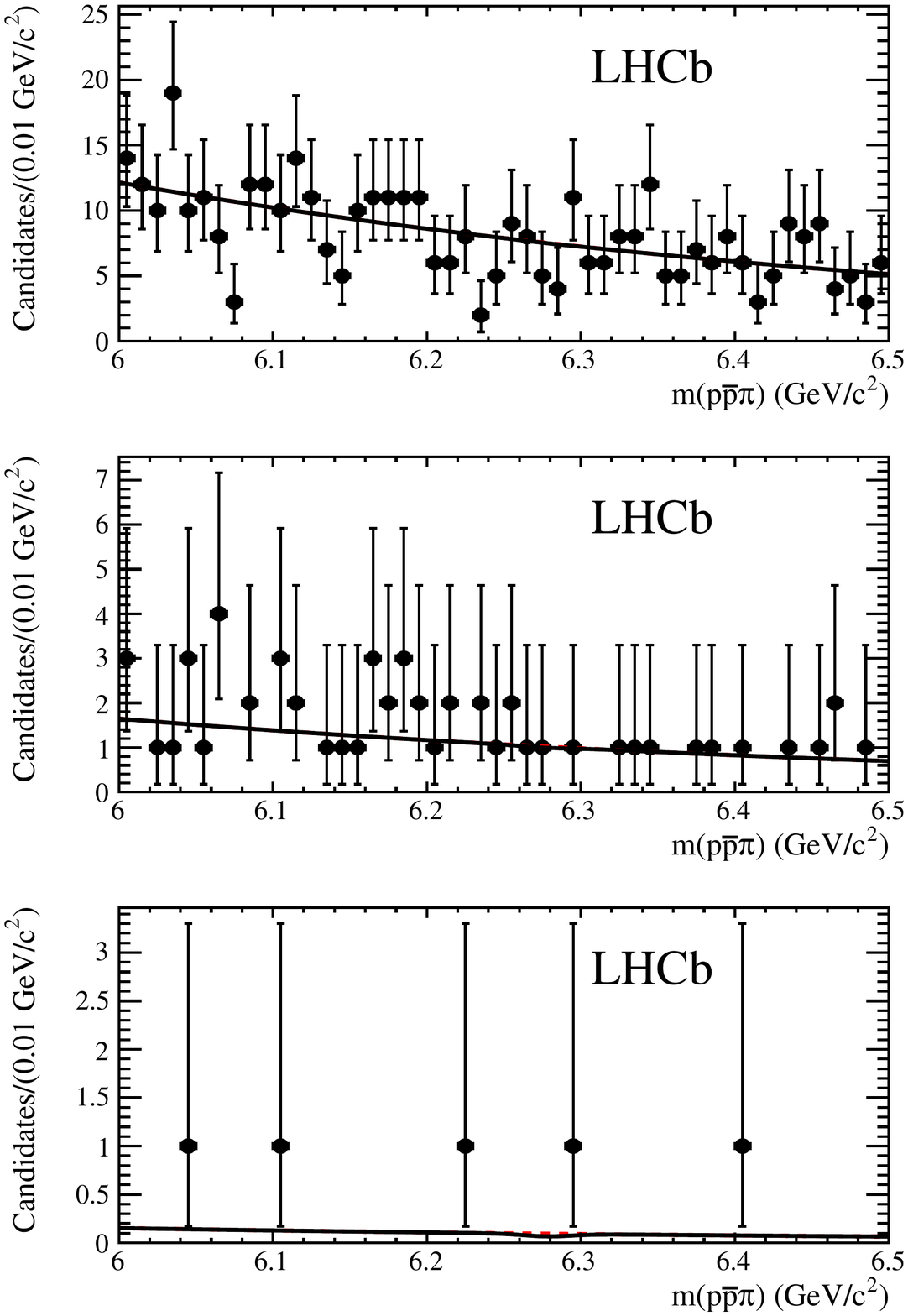}
\end{center}
\caption{Projection of fits to the $\proton\antiproton\pip$ invariant mass in the \Bc region, in the bins of BDT output (top) $0.04<X<0.12$, (middle) $0.12<X<0.18$ and (bottom) $X>0.18$, for (left) $m(\proton\antiproton)<2.85\gevcc$ and (right) $2.85<m(\proton\antiproton)<3.15\gevcc$. The red long-dashed lines represent the combinatorial background. The signal and partially reconstructed components are too small to be shown.}
\label{Fig:BcFits}
\end{figure*}

The simultaneous fits performed in the \Bc region are made for the region exclusive to the annihilation process, $m(\proton\antiproton)<2.85\gevcc$, and for the charmonium region, $2.85<m(\proton\antiproton)<3.15\gevcc$. The fraction of the yield of the partially reconstructed background in each bin of the BDT output is constrained to be the same as in the simulation. The results are shown in Fig.~\ref{Fig:BcFits}. The corresponding signal yields are $N(\bcpppiplus)=-2.7\pm6.3$ for $m(\proton\antiproton)<2.85\gevcc$ and $N(\bcpppiplus)=-0.1\pm3.0$ for $2.85<m(\proton\antiproton)<3.15\gevcc$.

The main observable under consideration is determined as
\begin{equation}
R_p \equiv \frac{f_c}{f_u}\times\mathcal{B}(\bcpppiplus)=\frac{N(\bcpppiplus)}{N(\pppiplus)}\times\frac{\epsilon_u}{\epsilon_c}\times\mathcal{B}(\pppiplus),
\end{equation}
and a cross-check is made for the \jpsi mode
\begin{equation}
R_p^\jpsi \equiv \frac{f_c}{f_u}\times\mathcal{B}(\Bc\to\jpsi\pip)=\frac{N(\Bc\to\jpsi(\to\proton\antiproton)\pip)}{N(\pppiplus)}\times\frac{\epsilon_u}{\epsilon_c^\jpsi}\times\frac{\mathcal{B}(\pppiplus)}{\mathcal{B}(\jpsi\to\proton\antiproton)},
\end{equation}
where the efficiencies $\epsilon$ are discussed in Sec.~5.

\section{Efficiencies}
\begin{figure*}[t]
\begin{center}
\includegraphics[width=0.45\textwidth]{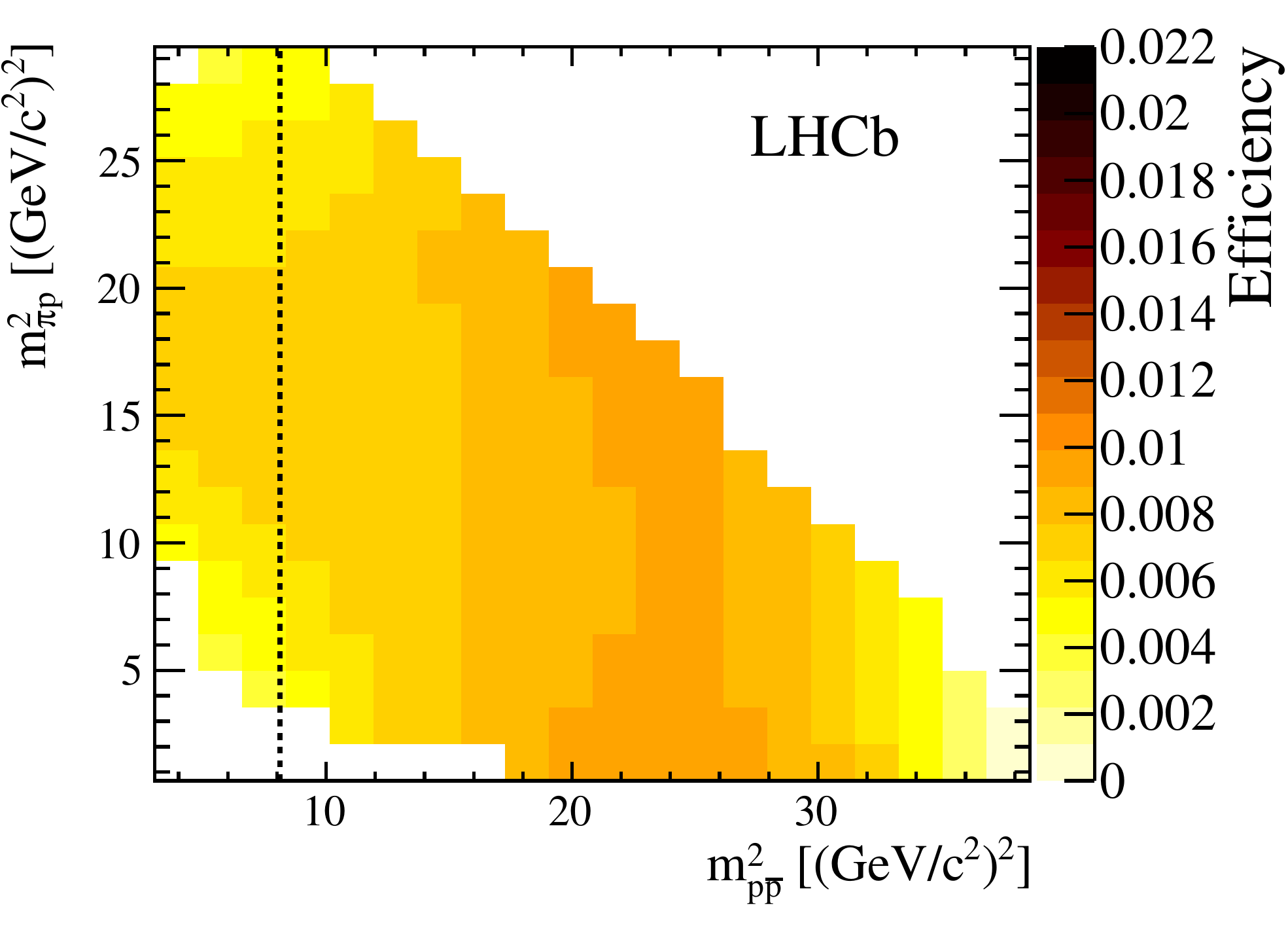}
\includegraphics[width=0.45\textwidth]{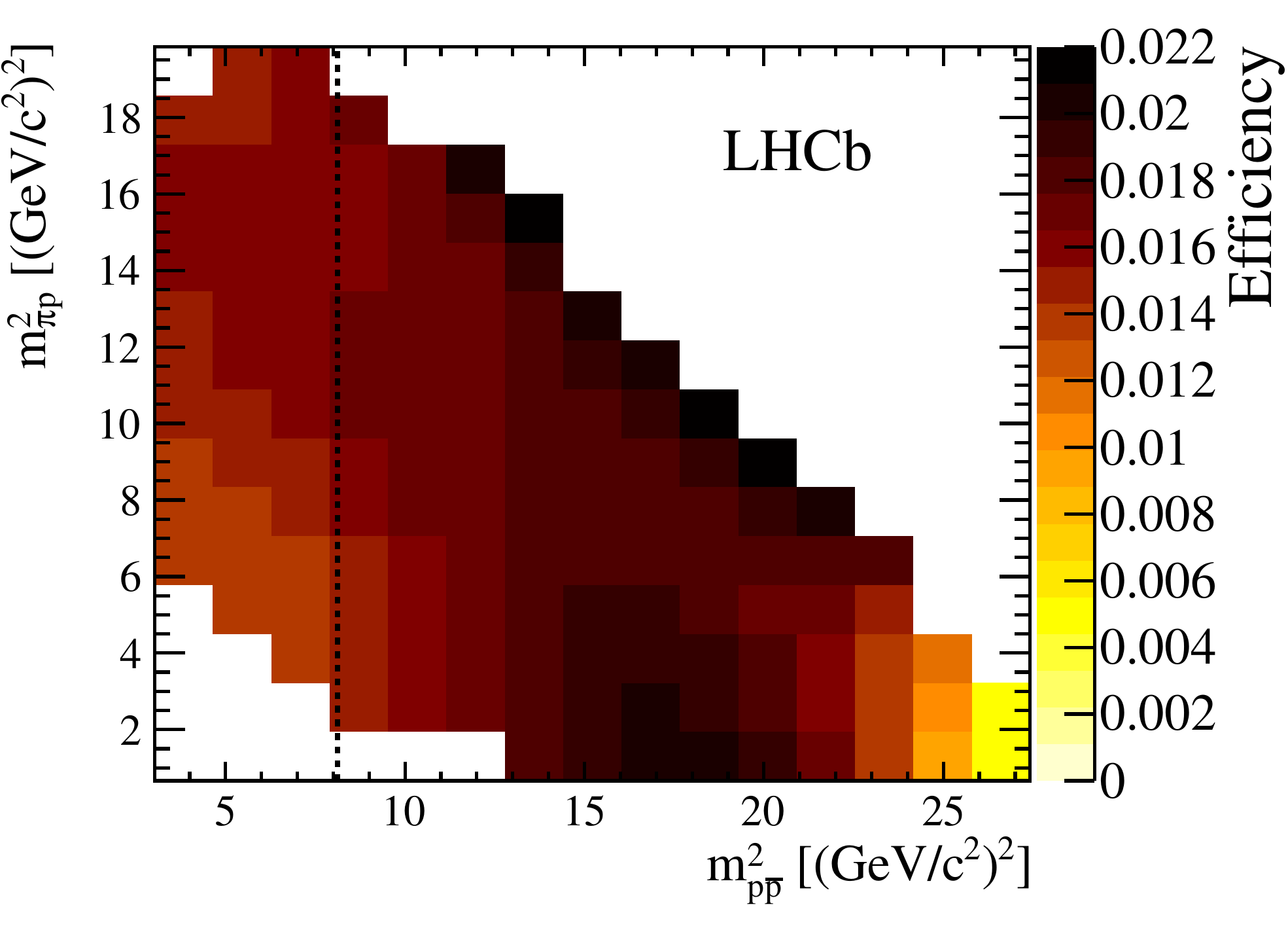}
\end{center}
\vspace{-0.5cm}
\caption{Combined acceptance in the plane $(m^2(\proton\antiproton),m^2(\proton\pi))$ for (left) \bcpppiplus and (right) \pppiplus events. The vertical dashed line corresponds to $m(\proton\antiproton)=2.85\gevcc$.}
\label{Fig:RecoAcc}
\end{figure*}

The reconstruction and selection efficiencies are computed from acceptance maps defined in the $m^2(\proton\antiproton)$ \vs $m^2(\proton\pi)$ plane. These maps include the effects of event reconstruction, triggers, preselection, BDT and PID selections, and are obtained from simulation for both \bcpppiplus and \pppiplus. The PID map is obtained by studying data-driven responses from calibration data samples of kinematically identified pions, kaons and protons originating from the decays $D^{*+}\to D^0(\to K^-\pi^+)\pi^+$, $\Lz\to p\pi^-$ and $\Lc\to p\Km\pip$. The maps are smoothed using fits involving two-dimensional fourth-order polynomials. Figure~\ref{Fig:RecoAcc} shows the final combination of these maps. 

To infer the average efficiency for \pppiplus, signal weights are calculated with the {\it sPlot} technique \cite{Pivk:2004ty} from the fits shown in Fig.~\ref{Fig:BuFits}. A weight is associated with each candidate depending on its position in the $m^2(\proton\antiproton)$ \vs $m^2(\proton\pi)$ plane. The acceptance maps are then used to determine an averaged efficiency, $\epsilon_u^{\mathrm{sel}}\equiv\langle \epsilon^{\mathrm{sel}}(\pppiplus) \rangle$. For \bcpppiplus, since no signal is available in data, a simple average is performed in the region $m(\proton\antiproton)<2.85\gevcc$ to obtain $\epsilon_c^{\mathrm{sel}}$, which leads to a substantial systematic uncertainty due to the variation of the efficiency over this region. 

In computing the ratio $\epsilon_u^{\mathrm{sel}}/\epsilon_c^{\mathrm{sel}}$, three corrections are needed to account for data-simulation discrepancies: tracking efficiency, hardware hadron trigger efficiency; and the fiducial region cuts $\pt(B)<20\gevc$ and $2.0<y(B)<4.5$. After these corrections, $\epsilon_u^{\mathrm{sel}}/\epsilon_c^{\mathrm{sel}}=2.495\pm0.028$ is obtained including associated systematic uncertainties.

Another efficiency ratio accounts for the fact that \pppiplus and \bcpppiplus decays are only detected if all the decay daughters are in the \lhcb acceptance: the fractions of events satisfying this requirement are estimated by simulation and are found to be $\epsilon_u^{\mathrm{acc}}=(18.91\pm0.10)\%$ and $\epsilon_c^{\mathrm{acc}}=(15.82\pm0.03)\%$, which gives $\epsilon_u^{\mathrm{acc}}/\epsilon_c^{\mathrm{acc}}=1.195\pm0.007$.

For $\Bc\to\jpsi(\proton\antiproton)\pip$, a similar procedure is applied and the following values are found: $\epsilon_u^{\mathrm{sel}}/\epsilon_c^{\jpsi,\mathrm{sel}}=2.513\pm0.032$ and $\epsilon_u^{\mathrm{acc}}/\epsilon_c^{\jpsi,\mathrm{acc}}=1.186\pm0.007$. The efficiency ratio used for the final results is $\epsilon_u/\epsilon_c=\epsilon_u^{\mathrm{sel}}/\epsilon_c^{\mathrm{sel}}\times\epsilon_u^{\mathrm{acc}}/\epsilon_c^{\mathrm{acc}}$. The differences between the \Bp and \Bc detector acceptance and selection efficiencies are caused by the different lifetimes and masses of the two mesons.

\section{Systematic uncertainties}

Part of the systematic uncertainties are related to the computation of the efficiency ratios, such as the PID calibration, the uncertainty in the \Bc lifetime, $0.507\pm0.009\ps$ \cite{HFAG}, the limited sizes of the simulation samples, the effect of the detector acceptance, the distribution of the BDT output, and the trigger and fiducial cut corrections.  
Others are related to the branching fractions $\mathcal{B}(\pppi)=(1.07\pm0.16)\times 10^{-6}$ \cite{LHCb-PAPER-2014-034} and $\mathcal{B}(\jpsi\to\proton\antiproton)=(2.120\pm0.029)\times 10^{-3}$ \cite{PDG2014}, or to the variation of the selection efficiency of \bcpppi over the phase-space region $m(\proton\antiproton)<2.85\gevcc$, due to the lack of knowledge of the kinematics in the absence of signal in data (modelling).

Table \ref{Tab:systematics} lists the different sources of systematic uncertainties. The PID uncertainty is dominated by the finite size of the proton calibration samples, which limits the sampling of the identification efficiency as a function of the track momentum and rapidity. A similar comment applies for the hardware trigger efficiency correction, where the effect is smaller due to a one-dimensional sampling as a function of the transverse momentum \pt. The uncertainty related to the differences in the BDT output shape between data and simulation has been estimated using $\Bp\to\proton\antiproton h^+$ ($h=K,\pi$) samples where the signal yield has been studied as a function of the requirements on the BDT output in both data and simulation. The uncertainty on the fit model, including the knowledge of the signal shape and the contribution of the partially reconstructed background, is found to have no impact on the final result.
\begin{table}[tb]
	\caption{Relative systematic uncertainties (in \%) on the ratio $\epsilon_u/\epsilon_c$ and input branching fractions.}
    \label{Tab:systematics}
  \begin{center}
\begin{tabular}{ccc}
\centering
 Source & $\bcpppiplus$, $m(\proton\antiproton)<2.85\gevcc$ & $\Bc\to\jpsi(\to\proton\antiproton)\pip$\\ 
\hline 
PID & 3.0 & 3.0\\
\Bc lifetime & 2.0 & 2.0\\
Simulation & 0.8 & 0.9\\
Detector acceptance & 0.6 & 0.6\\
BDT shape & 1.5 & 1.5\\
Hardware trigger correction & 0.8& 0.9\\
Fiducial cut & 0.1 & 0.1\\
Modelling & 15 & ---\\
$\mathcal{B}(\pppiplus)$ & 15 & 15\\
$\mathcal{B}(\jpsi\to\proton\antiproton)$ & --- & 1.4\\
\end{tabular}
\end{center}
\end{table}

\section{Results and summary}
Upper limits on $R_p$ and $R_p^\jpsi$ are estimated by making scans of these quantities, comparing profile likelihood ratios for the ``signal+background'' against ``background''-only hypotheses \cite{CL-art}. From these fits, $p$-value profiles are inferred, the signal $p$-value being the ratio of the ``signal+background'' and ``background'' $p$-values. The point at which the $p$-value falls below 5\% determines the 95\% confidence level (CL) upper limit. In the determination of this value, the systematic uncertainties, shown in Table \ref{Tab:systematics}, and the statistical uncertainty on the normalization channel yield are taken into account.

\begin{figure*}[t]
\begin{center}
\includegraphics[width=0.45\textwidth]{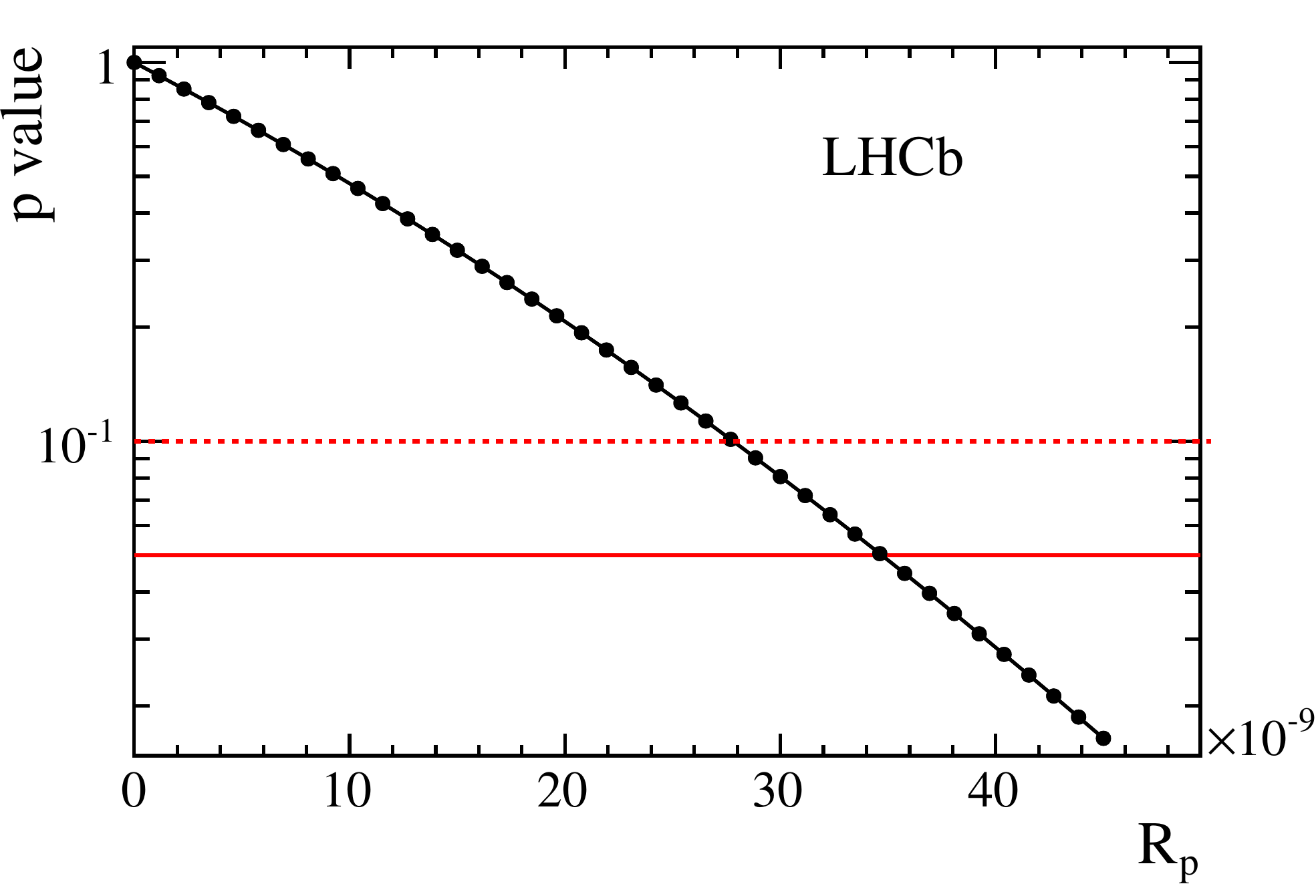}
\includegraphics[width=0.45\textwidth]{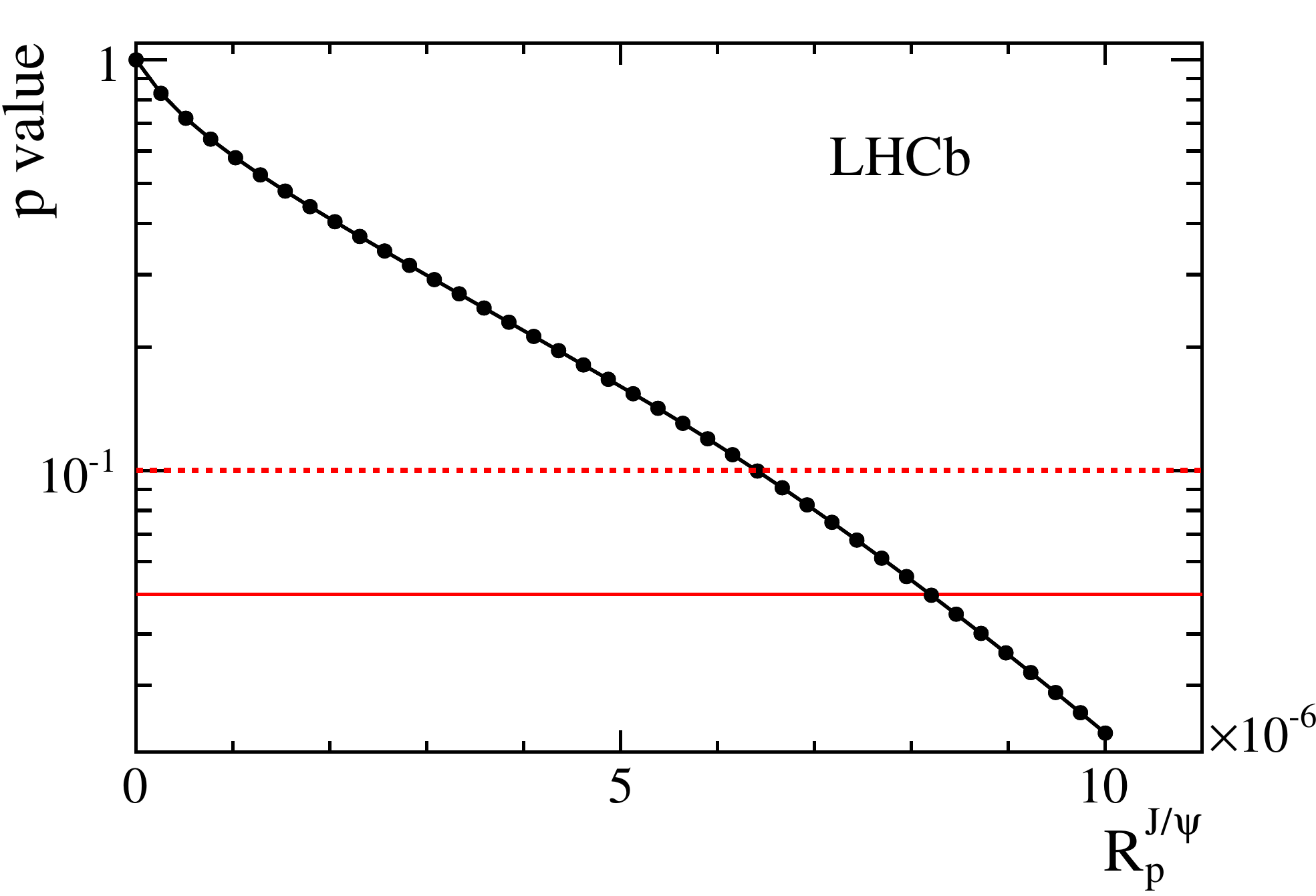}
\end{center}
\vspace{-0.5cm}
\caption{$p$-value profile for (left) $R_p$ and (right) $R_p^\jpsi$. The horizontal red solid and dashed lines indicate the 5\% and 10\% confidence levels.}
\label{Fig:pvalue}
\end{figure*}

The $p$-value scans are shown in Fig.~\ref{Fig:pvalue}, from which the following values are found: $R_p<3.6\times10^{-8}$ ($m(\proton\antiproton)<2.85\gevcc$) and $R_p^{\jpsi}<8.4\times10^{-6}$ at 95\% CL. The latter limit is compatible with a measurement of $\frac{f_c}{f_u}\times\frac{\mathcal{B}(\Bc\to\jpsi\pip)}{\mathcal{B}(\Bu\to\jpsi\Kp)}$ \cite{LHCb-PAPER-2014-050} from which the value $R_p^{\jpsi}=(7.0\pm0.3)\times10^{-6}$ is inferred. At 90\% CL, the limits are $R_p<2.8\times10^{-8}$ and $R_p^{\jpsi}<6.5\times10^{-6}$.

In summary, a search for the $\overline bc$ annihilation process leading to \Bc meson decays into the $\proton\antiproton\pip$ final state has been performed for the fiducial region $m(\proton\antiproton)<2.85\gevcc$, $\pt(B)<20\gevc$ and $2.0<y(B)<4.5$. No signal is observed and a 95\% confidence level upper limit is inferred, 
\begin{center}
$R_p = \frac{f_c}{f_u}\times\mathcal{B}(\bcpppiplus)<3.6\times10^{-8}$.
\end{center}








\section*{Acknowledgements}

\noindent We express our gratitude to our colleagues in the CERN
accelerator departments for the excellent performance of the LHC. We
thank the technical and administrative staff at the LHCb
institutes. We acknowledge support from CERN and from the national
agencies: CAPES, CNPq, FAPERJ and FINEP (Brazil); NSFC (China);
CNRS/IN2P3 (France); BMBF, DFG and MPG (Germany); INFN (Italy); 
FOM and NWO (The Netherlands); MNiSW and NCN (Poland); MEN/IFA (Romania); 
MinES and FANO (Russia); MinECo (Spain); SNSF and SER (Switzerland); 
NASU (Ukraine); STFC (United Kingdom); NSF (USA).
We acknowledge the computing resources that are provided by CERN, IN2P3 (France), KIT and DESY (Germany), INFN (Italy), SURF (The Netherlands), PIC (Spain), GridPP (United Kingdom), RRCKI and Yandex LLC (Russia), CSCS (Switzerland), IFIN-HH (Romania), CBPF (Brazil), PL-GRID (Poland) and OSC (USA). We are indebted to the communities behind the multiple open 
source software packages on which we depend.
Individual groups or members have received support from AvH Foundation (Germany),
EPLANET, Marie Sk\l{}odowska-Curie Actions and ERC (European Union), 
Conseil G\'{e}n\'{e}ral de Haute-Savoie, Labex ENIGMASS and OCEVU, 
R\'{e}gion Auvergne (France), RFBR and Yandex LLC (Russia), GVA, XuntaGal and GENCAT (Spain), Herchel Smith Fund, The Royal Society, Royal Commission for the Exhibition of 1851 and the Leverhulme Trust (United Kingdom).


\ifx\mcitethebibliography\mciteundefinedmacro
\PackageError{LHCb.bst}{mciteplus.sty has not been loaded}
{This bibstyle requires the use of the mciteplus package.}\fi
\providecommand{\href}[2]{#2}

\newpage
\centerline{\large\bf LHCb collaboration}
\begin{flushleft}
\small
R.~Aaij$^{39}$, 
C.~Abell\'{a}n~Beteta$^{41}$, 
B.~Adeva$^{38}$, 
M.~Adinolfi$^{47}$, 
Z.~Ajaltouni$^{5}$, 
S.~Akar$^{6}$, 
J.~Albrecht$^{10}$, 
F.~Alessio$^{39}$, 
M.~Alexander$^{52}$, 
S.~Ali$^{42}$
% ...

{\footnotesize \it
% ...
$ ^{5}$Clermont Universit\'{e}, Universit\'{e} Blaise Pascal, CNRS/IN2P3, LPC, Clermont-Ferrand, France\\
$ ^{6}$CPPM, Aix-Marseille Universit\'{e}, CNRS/IN2P3, Marseille, France\\
% ...
$ ^{10}$Fakult\"{a}t Physik, Technische Universit\"{a}t Dortmund, Dortmund, Germany\\
% ...
$ ^{38}$Universidad de Santiago de Compostela, Santiago de Compostela, Spain\\
$ ^{39}$European Organization for Nuclear Research (CERN), Geneva, Switzerland\\
% ...
$ ^{41}$Physik-Institut, Universit\"{a}t Z\"{u}rich, Z\"{u}rich, Switzerland\\
$ ^{42}$Nikhef National Institute for Subatomic Physics, Amsterdam, The Netherlands\\
% ...
$ ^{47}$H.H. Wills Physics Laboratory, University of Bristol, Bristol, United Kingdom\\
% ...
$ ^{52}$School of Physics and Astronomy, University of Glasgow, Glasgow, United Kingdom\\
% ...
}
\end{flushleft}

\end{document}